# Algebraic Form of Malliavin Calculus: Creation-Annihilation Operators, Conserved Currents and All That

Peter B. Lerner[1]

## Abstract

The extremely useful method of Malliavin calculus has not yet gained adequate popularity because of the complicated analytical apparatus of this method. The author attempts here to propose a simplified algebraic formalism similar to Malliavin calculus, but based on the notion of creation-annihilation operators instead of Malliavin derivative to replace analytical theorems with algebraic computations. Three test problems: the valuation of portfolio with stochastic payoff function, the expression of the terminal payoff through stochastic integral and the approximate equation for the high-frequency market measure are discussed in Appendices.

## Contents

---

1   SciTech, LLC, Woodland Dr., State College, PA 16803, pblerner@syr.edu, pbl2@psu.edu.

**Introduction**

Malliavin calculus was formulated by Malliavin and co-authors in the late 1970s (see Malliavin and Thalmaier [Ma1], for bibliography up to 2005). Yet, this extremely useful technique has not gained popularity within the finance community because of analytical difficulties involved in this extension of stochastic calculus. In particular, only a limited number of papers on ssrn.com (a fair sample of research in mathematical finance) use Malliavin calculus and the list of authors includes mostly investigators trained as professional mathematicians. Furthermore, the main extant financial application of Malliavin calculus—the computation of greeks—becomes less and less pertinent with the fast growth of computational power and the possibility of direct reanalysis of the problem instead of using elasticities calculated through greeks.

There was a single attempt to infuse stochastic calculus with concepts based on physical intuition by V. N. Kolokoltsov [Kol1, Kol2] but this attempt was also little noticed outside of the mathematics circles. Wherever similar ideas appeared under the guise of "Quantum Stochastic Calculus" beginning with the classical paper by Belyavkin [Bel1], they were developed to enrich quantum physics more than finance.

In this paper, I propose a simple set of algebraic rules that allow computations similar to Malliavin calculus without reference to the analytical apparatus unknown or forgotten by the majority of people educated in conventional physics or finance programs. The author does not claim that his technique is formally equivalent to the "full" Malliavin calculus. The mathematically rigorous definition of the creation-annihilation operators in the context of stochastic calculus were provided, e.g. by Hudson and Parthasarathy [Hud1], Belyavkin [Bel2], Kolokoltsov [Kol2] and *op. cit.* on the basis of the Fock space and Wick ordering. Yet, Belyavkin's approach is based on Schrödinger's equation for stochastic environments. In many formulations, though, using Heisenberg or interaction representations seems more natural.

To justify my effort with professionals, I emphasize that simplifications and popularizations have a venerable history in mathematics. The Wiener measure and Wiener integral were almost unknown outside of mathematical circles until Feynman produced now-celebrated Feynman-Kac formula for the Schrödinger equation (with a naïve complexification $t \to i\tau$, [Fey1]). The theory of Lie groups was considered almost impenetrable for teaching to students before Dynkin invented his diagrams as a teaching implement for presenting the Van Der Warden theorem at a seminar. The Dirac's delta function found almost immediate applications in physics and electrical engineering, yet despite rigorous mathematical foundations being available since the 1930s from the works of Sobolev [Sobol1, Maz1], the generalized functions or distributions did not achieve popularity before the development of an alternative definition around 1950 and subsequent books by Laurent Schwarz. [Schwarz1] Obviously, in the realms of mathematics and physics, the belated realization that "que je dis de la prose sans que j'en susse rien"[2] possesses some additional benefits.

Any formal structure deserving the name of "calculus" has to possess several properties. These usually are:

- The operation of differentiation.
- A possibility to define integration, which is in a certain sense a reverse operation to differentiation (e.g. Skorokhod integral for Malliavin calculus).
- Differentiation can be interpreted as a linear part of the change of a function (functional, operator, etc.). This property is reflected in the Lagrange (mean value) theorem of elementary calculus or the Clark-Ocone theorem in stochastic calculus (Lagrange, Clark-Ocone).
- Usefulness of the defined operation of differentiation in the solution of optimization problems (e.g. maximum and minimum theorems in elementary calculus [Ru2] or Hamilton-Jacobi-Bellman (HJB) equation in optimization and stochastic calculus [Trout1]).
- The possibility of interpreting dynamic equations in terms of (hydrodynamic) flows (for instance, Gauss theorems in vector calculus [Galb1].
- Demonstration of the usefulness of proposed formalism in treatment of some maybe already solved problems by this new method.

Algebraization of stochastic calculus is performed using creation-annihilation operators with conventional bosonic rules of operations, which can be extracted from most textbooks of quantum mechanics (further QM, see, e.g. [Cohen1], [Schw1], [Sh1]).[3] Below, I summarize these rules.

---

2   J.-B. Moliere, *Bourgeois Gentilhomme*.
3 Most of the QM textbooks for historical reasons—the students are supposed to follow up with scattering theory, etc.—begin with the Schrödinger equation and later introduce creation and annihilation operators within the formalism of secondary quantization, which is usually left to textbooks on many-body theory. ([Sak1], [Sh1]) Only recently, with the proliferation of studies in quantum optics, has the approach to QM starting with the Heisenberg operators returned (e.g. [Mil1]).

The presented algebraic formalism leaves open a question: what significance, if any, might be ascribed to the fermion sector? In quantum mechanics, the fermion sector not only appears naturally but also can be united in supersymmetric structures with the boson sector of almost any QM problem. In particular, for the supersymmetric representation of multiphoton transitions—or the higher-order polynomial Hamiltonians—one may consult the author's work [Andr1]).

Boson creation-annihilation operators act on elements of a separable Hilbert space $L$=L$^2$(C). The elements of $L$ can be identified with complex square integrable functions. As usual, one starts with the momentum and coordinate operators defined according to the rules:

$$\hat{q}\,\psi \rightarrow q\,\psi$$

$$\hat{p}\,\psi \rightarrow -i\,\nabla_q\,\psi$$

(1)

where $\psi$ is an element of $L$, i.e., a square integrable martingale in the parlance of probability theory. Note that the momentum operator is defined on a (dense) subspace of L. This construction is obviously heuristic.

However, there are efforts to provide a rigorous foundation of this construction on the basis of the Fock space [Acc1, Apl2, Belt1, Kon1] and Gelfand triples or rigged Hilbert spaces ([Gel1], Chapters 2-4, [Mad1] and *op. cit.,* [Ob1]). The definition of coordinate and momentum operators by Equation (1) is obviously correct for the Euclidean configuration space of arbitrary dimension. However, for simplicity, we shall use one-dimensional configuration space in most of our treatment and refer to multi-dimensional generalizations only as required.

§1. **Creation and annihilation** operators are defined by the standard equalities:

$$a = \frac{\hat{q}+i\,\hat{p}}{\sqrt{2}} \equiv \frac{q+\partial_q}{\sqrt{2}}$$

(2)

$$a^+ = \frac{\hat{q}-i\,\hat{p}}{\sqrt{2}} \equiv \frac{q-\partial_q}{\sqrt{2}}$$

Momentum and coordinate operators are expressed through creation-annihilation operators as follows:

$$\hat{q} = \frac{a^+ + a}{\sqrt{2}}$$

(2')

$$\hat{p} = \frac{a - a^+}{\sqrt{2}\,i}$$

Creation and annihilation operators obey canonical commutation relations for bosons:

$$a\,a^+ - a^+\,a \equiv [a, a^+] = 1$$

(3)

<u>Note 1</u>: Compare this to Malliavin and Thalmaier (Ma1, 2005, Chapter 1).[4]

<u>Note 2</u>: For multi-dimensional space, commutation relations of Equation (3) acquire the following form:

$$a_k\,a_{k'}^+ - a_{k'}^+\,a_k = \delta_{kk'}$$

(3')

where $k$ and $k'$ are discrete or continuous indexes and $\delta$ is a Kronecker symbol or delta function, respectively. In a probabilistic sense, where evolution of the $\vec{q}$ is a standard diffusion, Equation (3') reflects the mutual independence of the Brownian motions corresponding to each Euclidean coordinate. Furthermore, we can expect that the "deformed" commutation relationships typical for the quantum groups:

$$a_k\,a_{k'}^+ - q\,a_k^+\,a_{k'} = \delta_{kk'}$$

(3'')

correspond to the correlated Brownian motions. Derivation of Robertson-Schrödinger [Rob1, Sch1] uncertainty relations from Equation (3'') is left as an exercise.

§ 2. **The Fock space** in stochastic context. One of advantages of the method of second quantization in

---

4    Mailliavin and Thalmaier ([Ma1], Section 1.1) define creation operator as $x$ and annihilation operator as $x\,\partial_x - 1$ with the same canonical commutation relation but never use this construction elsewhere in the book except for Appendix A, where the authors redefine them again as Fourier components of the field. Note that, in contrast to the conventional QM definition, Ma1 defines creation and annihilation operators in Chapter 1 as self-adjoint.

quantum mechanics is the possibility to describe an operator algebra on a Hilbert space by a "Hilbert" space of its own, the Fock space. ([Cohen1], [Schw1])

In Malliavin and Thalmaier, the state space of the problem is $L^2(W, \mathrm{d}x)$ where $W$ is a symbolic notation for the space of continuous trajectories between 0 and 1 with $f \in W$, $f(0)=0$. This space is topologically complicated.

Elementary construction of the Fock space can be described as follows. First, one selects a countable orthonormal basis in a Hilbert space $L^2(C)$. Heisenberg creation and annihilation operators act on the basis functions $|e_n>$ from the state space as follows:

$$a\,|e_n> = \sqrt{n}\,|e_{n-1}>$$
$$a^+|e_n> = \sqrt{n+1}\,|e_{n+1}>$$

(4)

As basis vectors for $L^2$ it is conventional to take Hermit polynomials $H_n$ (harmonic oscillator basis). Then, loosely speaking, the action of any polynomial of the creation and annihilation operators on a state vector can be expressed as a map from $L^2(C)$ to its $n$th power, where $n=p-q$, $p$—leading degree of creation operators, and $q$—leading degree of annihilation operators in the polynomial.

To represent a wide class of stochastic processes in this convenient form, we can use the theorem ([Kar1], Sect. 3.4 and [Øk2], Sect. 4.2 and 12.3) on the representation of square-integrable martingales as Wiener integrals.[5] Namely, a square-integrable martingale can be represented as a Wiener integral with a predictable process $f$ as an integrand:

$$X_t = \int_0^t f_u\, dW_u$$

(5)

$$|e_n>(x) = \frac{1}{\sqrt{n!}} \int_0^x H_n\left(\frac{u}{\sqrt{2}}\right) e^{\frac{-u^2}{2}}\, dW_u$$

(6)

A central role in theory is played by a self-adjoint function of $a$ and $a^+$, which is called Hamiltonian $H(a,a^+)$. For now, we shall consider Hamiltonians polynomial functions of the creation and annihilation operators but later we shall discuss reasonable extensions of this class. The simplest Hamiltonian (called a one-dimensional harmonic oscillator in QM) can be obtained by adding the Hermitian conjugate to the simplest binomial:

$$a^+ a \rightarrow \frac{(a^+ a + a\, a^+)}{2} \equiv a^+ a + \frac{1}{2}$$

---

5  Equation (7) below represents a square-integrable martingale. This formulation might be too narrow even for the infinitely divisible processes, many of which are not square-integrable. In that context, requirement of square integrability in theorems of the Chapters 3-4 in [Øk1] seems as very restrictive.

Time evolution of the basis vector under a "free" Hamiltonian, $H_0 = \omega\left(a^+ a + \frac{1}{2}\right)$ is harmonic, i.e. it is represented by with Euclidean time change $t \rightarrow i\tau$ by the formula:

$$|e_n > (x, \tau) = \frac{e^{-\omega(n+\frac{1}{2})\tau}}{\sqrt{n!}} \int_0^x H_n\left(\frac{u}{\sqrt{2}}\right) e^{\frac{-u^2}{2}} \, d\, W_u \qquad (7)$$

The connection of the Equation (7) with the Ornstein-Uhlenbeck process is clear from the ability to represent time evolution of an arbitrary process belonging to $L^2$ through the following equalities. Namely, if

$$X_0 \sim \sum c_n |e_n > (x) \qquad (8)$$

then

$$X_t = e^{-\tau H_0} X_0 \sim \sum c_n e^{-\omega(n+\frac{1}{2})\tau} |e_n > (x)$$

§3. **Operators form an algebra C\*** with three additional operations: commutator of operators [ . , . ], Hermitian conjugation ($a \rightarrow a^+,\ a^+ \rightarrow a,\quad U \rightarrow \bar{U}^T$ ) where $U$ is any complex-valued numerical matrix (in particular $i\,c \rightarrow -i\,\bar{c}$ for any complex number) and normal (anti-normal) ordering. Furthermore, polynomial functions of the creation and annihilation operators are well-defined. Normal (anti-normal) ordering of the element $F(a, a^+)$ from the algebra is the ordering of of all creation operators to the left (right) of all annihilation operators using Equation (3).[6]

---

§4. **δ-Derivative**. Let $\acute{M}\left(a^+,a\right)$ and $\grave{M}\left(a,a^+\right)$ be normally and anti-normally ordered copies of the C* monomial:

$$\acute{M}\left(a^+,a\right)=\left(a^+\right)^m a^n$$

$$\grave{M}\left(a,a^+\right)=a^n\left(a^+\right)^m \tag{9}$$

with $m$ and $n$ arbitrary non-negative integers. Then, the derivatives in operator algebra C* can be defined by the following equalities:

$$\frac{\delta\acute{M}}{\delta a}=n\left(a^+\right)^m a^{n-1}$$

$$\frac{\delta\grave{M}}{\delta a^+}=m a^n\left(a^+\right)^{m-1} \tag{10}$$

The derivatives of constants are assumed to be zero. By the algebra properties of C* these definitions can be extended for an arbitrary polynomial member of C*.

§5. **Integration in C*** is defined as a reverse operation with respect to differentiation. By definition, the integrals of monomials from Equation (5) are:

$$\int\acute{M}\left(a,a^+\right)da=\frac{1}{n+1}\left(a^+\right)^m a^{n+1}$$

$$\int\grave{M}\left(a,a^+\right)da^+=\frac{1}{m+1}a^n\left(a^+\right)^m \tag{11}$$

<u>Note 1</u>: We presume that:

$$\int\dot{a}\,dt=\int d\,a$$
$$\int\dot{a^+}\,dt=\int d\,a^+$$

These equalities can be proven from Equations (17), §6, if one takes a Hamiltonian of the simplest form $H = a^+ a$.

Note 2: The following formal integral is equal to unity

$$\int a\, d\, a^+ - \int a^+ da = 1$$

as a consequence of canonical commutation relations.

Note 3: For the polynomial functions of creation and annihilation operators, the above defined integral is obviously inverse to the operation of δ-differentiation. So, in this context, the integration operation coincides with the Skorohod integral. [Sk1]

§6. **A Hilbert Space from Abstract Creation-Annihilation Operators.** As it is well known from Gelfand-Naimark -Segal theorems, GNS [Dix1, Kad1], every abstract operator algebra under some generic conditions can be realized as a Banach algebra on a Hilbert space. However, this realization is not constructive in the sense that a positive functional in the construction of GNS is not always given. In our case, where we postulated the Hamiltonian (§5), we can use a Gibbs measure to define a positive functionals on the operator space:

$$E_G[.] \equiv \int . \, d\mu_H = Tr(.\rho_z) \tag{12}$$

with

$$\rho_z(\hat{H}) = \frac{\exp^{-\beta\hat{H}}}{Z} \quad \text{and} \quad Z = Tr(e^{-\beta\hat{H}}), \beta > 0$$

and

$$\langle A, B \rangle = E_G[A B^+] \tag{13}$$

In Equation (12), operator trace can be defined in any suitable way (for instance, as a Diximier trace). Note, that if the spectrum of $H$ is non-degenerate and limited from above, the product (13) has a trivial zero subspace. Henceforth, Gibbs' measure induces a scalar product on the operator algebra itself. Otherwise, we have to choose a zero subspace of the expectation according to the Gibbs measure and proceed with the GNS as usual.

§7. **Heisenberg equations**. Further definitions that are needed to clarify the optimization properties of derivative, similar to the Euler-Lagrange equation, require dynamic description. To define dynamics, one considers a family of states (elements of an original Hilbert space) indexed by real parameter t: $\phi(t)$ normalized on unity: $\langle \phi(t), \phi(t) \rangle = 1$. Under some generic conditions (see [Ru1]) one can consider this family as being generated by an action of a unitary semigroup from a single element:

$$\phi(t) = e^{it\hat{H}} \phi(0) \equiv \hat{U}_t \phi(0) \tag{14}$$

In Equation (12), $\hat{H}$ is a self-adjoint operator, which can be further identified with the Hamiltonian. This construct is well known from Quantum Mechanics under the name of Heisenberg representation (see, e.g. [San1], [Schw1]). Dirichlet forms involving $\phi(t)$ are invariant, if one replaces the action of a semigroup on the element of the state space by a unitary transformation of an operator:

$$\langle \phi(t), \hat{A}\phi(t) \rangle = \langle \phi(0), \hat{A}(t)\phi(0) \rangle \tag{15}$$

where $\hat{A}$ is an operator acting on $L$ and

$$\hat{A}(t) = e^{-it\hat{H}} \hat{A} e^{it\hat{H}} \tag{16}$$

Then, applying Baker-Hausdorff formula:

$$e^{\hat{A}+\hat{B}} = e^{\hat{A}} e^{\hat{B}} e^{[\hat{A},\hat{B}]} \tag{17}$$

immediately results in the Heisenberg Equation (equation of motion for the operator $\hat{A}(t)$ ) where the dot signifies time derivative:

$$\dot{\hat{A}} = i[\hat{H}, \hat{A}] \tag{18}$$

In particular:

$$\dot{a}(t) = -[\hat{H}, a] \tag{19}$$

$$\dot{a}^+ = [\hat{H}, a^+]$$

<u>Note 1</u>: Commutator in our approach is analogous to the Lie bracket in Malliavin calculus (Malliavin and Thalmaier, 2005, Section 5.3).

<u>Note 2</u>: We shall use time as it is used in quantum mechanics to highlight the unitary nature of evolution operators. Only for the applications where it is essential, shall we return to the Euclidean convention of $t \to i\tau$ used in statistical physics and mathematical finance .

§8. **Schrodinger equations**. With the help of the δ operation of differentiation (§3), the Heisenberg Equations (18) for the creation-annihilation operators can be rendered in Hamilton form, which superficially resembles Schrodinger equations:

$$i \frac{da}{dt} = \frac{\delta \hat{H}(a, a^+)}{\delta a}$$

(18)

$$-i \frac{da^+}{dt} = \frac{\delta \hat{H}(a, a^+)}{\delta a^+}$$

<u>Note</u>: The integral defined in §5 allows for a formal integration of Equations (18).

§9. **Law of energy conservation**. Because the Hamiltonian operator (we deliberately consider time-independent Hamiltonians) commutes with itself

$$[\hat{H}, \hat{H}] = 0$$

(19)

the Hamiltonian is an integral of motion:

$$\langle \phi, \hat{H}\phi \rangle = const = E$$

(20)

The law of energy conservation is consistent with the δ-derivative. Indeed:

$$\partial_t \langle \phi, \hat{H}\phi \rangle = \langle \phi, (\frac{\delta \hat{H}}{\delta a^+}\dot{a}^+ + \dot{a}\frac{\delta \hat{H}}{\delta a})\phi \rangle$$

$$= i \langle \phi, (\frac{\delta \hat{H}}{\delta a}\frac{\delta \hat{H}}{\delta a^+} - \frac{\delta \hat{H}}{\delta a}\frac{\delta \hat{H}}{\delta a^+})\phi \rangle = 0$$

(21)

where we used, firstly, Equations (18) and, secondly, the property that the Hamiltonian is self-adjoint (§2). As before, we assume that the Hamiltonian does not contain parameter $t$ explicitly and depends on time only through the Heisenberg operators $a$, $a^+$.

§10. **Euler-Lagrange equations**. To explore applications of differential calculus to optimization problems, we have to define action (time-indexed family of operators) and/or Lagrangian. Then we can formulate the analog of variation calculus either through the principle of minimum action or the Euler-Lagrange equations, which is our goal. The action can be defined by the conventional integral equality:

$$S = \int p\,dq - \int H\,dt \rightarrow \frac{1}{2i}\int (a - a^+)d(a + a^+) - \int H(a, a^+)dt$$

(22)

Using the note after §5 and the energy conservation principle (§9), we obtain the expression for action in the following form:

$$\hat{S}_t = \frac{1}{4i}(a^2 - a^{+2}) - Et + \frac{1}{2i}$$

(23)

As usual, a "momentum" operator can be obtained as (formal) coordinate derivative of the action:

$$\hat{p} = \frac{\partial \hat{\hat{S}}}{\partial a}\frac{\partial a}{\partial \hat{q}} + \frac{\partial \hat{\hat{S}}}{\partial a^+}\frac{\partial a^+}{\partial \hat{q}}$$

$$= \frac{\partial \hat{\hat{S}}}{\partial a}(\frac{\partial a}{\partial \hat{q}})^{-1} + \frac{\partial \hat{\hat{S}}}{\partial a^+}(\frac{\partial a^+}{\partial \hat{q}})^{-1} = \frac{\partial \hat{\hat{S}}}{\partial \hat{q}}$$

(24)

where we use the expressions of Equations (2) and (2').

§11. **Lagrangian density**. Lagrangian (more accurately, Lagrangian density[7]) is obtained, as is usual in classical mechanics ([Mo1], [Tay1]), as a time derivative of action:

$$L = \frac{1}{2i}\left(a\dot{a} - a^+ \dot{a}^+\right) - \hat{H}\left(a, a^+\right)$$

(25)

or

$$L = \frac{1}{2}\left(\hat{p}\dot{\hat{q}} + \dot{\hat{q}}\hat{p}\right) - H\left(\hat{p}, \hat{q}\right)$$

(26)

In the Lagrangian approach, time derivatives of creation-annihilation operators should be considered as independent variables, rather than the products of Heisenberg transformation of original operators (§5). Because, by construction, the Hamiltonian operator does not depend on them, a conventional Euler operation applied to Lagrangian density results in:

$$\left(\frac{d}{dt}\frac{\delta}{\delta\dot{\hat{q}}} - \frac{\delta}{\delta\hat{q}}\right)L = 0$$

(27)

In (27), we consider Lagrangian as a function of independent variables $\dot{\hat{q}}, \hat{q}$. We define $q$-derivative as defined by conventional chain rules with respect to $a$, $a^+$ as we already did in derivation of Equation (24). Returning to the creation-annihilation operators, we get:

$$i\dot{a} = \frac{\delta\hat{H}}{\delta a}$$

$$i\dot{a}^+ = -\frac{\delta\hat{H}}{\delta a^+}$$

(28)

Equations (28) reproduce the Schrödinger form of Equations (17) as they should be. However, there is an additional benefit in deriving them this way because the derivation of §5 (by direct computation) formally applies only to the Hamitonians polynomial in operators $a$ and $a^+$. This "derivation"—in effect, we postulated that application of the Euler operation to the Lagrangian density is zero—applies to any formal function of creation-annihilation operators.

---

7   Formally, to obtain Lagrangian from (25), one needs to define the measure on the state space and integrate the expression of Equation (25) over this measure.

Note: Polynomial operator algebras seem to be a very limited setting for calculus, yet the only one for which rigorous mathematical foundations are obvious. My proposal for a practically adequate extension is as follows. We postulate that the symbol $f(a^+, a) \rightarrow f(\bar{\alpha}, \alpha)$ of an operator, where $\alpha, \bar{\alpha}$ are $c$-numbers and $f$ is a complex function, can be expressed as asymptotic series of its arguments (for symbols of operators see [Hör1]).

§12. **Hamilton-Jacobi equation**. To clarify further the extremal properties of the δ-derivative, we have to deduce a formal analog of the Hamilton-Jacobi equation. Taking full time derivative of the action (Equations (25-27)) and using Equations (28) for expressing time derivatives of the creation and annihilation operators, we obtain:

$$\frac{d}{dt}\hat{S}_t - \frac{1}{2}(\frac{\delta H}{\delta a} a + a^+ \frac{\delta H}{\delta a^+}) = -H(a, a^+) \tag{29}$$

To reduce Equation (29) to the conventional form:

$$\partial_t \hat{S}_t + H = 0 \tag{30}$$

we must admit the following identification:

$$\frac{d}{dt} \rightarrow \frac{\partial}{\partial t} + \frac{1}{2}(\frac{\delta H}{\delta a} a + a^+ \frac{\delta H}{\delta a^+}) \tag{31}$$

The heuristic meaning of the second term in Equation (31) is that it is similar to affine connection in the formula for covariant derivative:

$$\hat{V} = \frac{1}{2}(\frac{\delta H}{\delta a} a + a^+ \frac{\delta H}{\delta a^+}) \tag{32}$$

Operator $V$ is explicitly self-adjoint. By analogy with conventional differential geometry, this operator must have a geometric interpretation.[8]

§13. **Liouvillian**. One of the important elements in Malliavin calculus is the possibility to define divergence, which has geometric significance [Tha1]. However, the geometric interpretation of the Laplacian and harmonic functions in conventional stochastic calculus requires complicated analytical reasoning. [Kar1] Furthermore, the utility of divergence as a concept is intertwined with an ability to invoke a "hydrodynamic" interpretation of divergence in terms of the flow of incompressible liquid. ([Mo1]) The most useful consequence of the concept of divergence is the Gauss theorem of conventional analysis. [Galb1]

It seems that geometric interpretation is only possible if one imposes some kind of simplectic structure on the state space of the system, which requires duplication of the state space $L$. However, the mathematical training of this author is insufficient to push this consideration further.

We define a new, duplicated state space $F$:

$$F = \bar{L} \otimes L$$

(33)

We identify a second copy of the Hilbert space $F$ with the space of linear functionals on $L$ ($L^2$-type space) by the elementary theorem from functional analysis [Ru1].

The elements of $F$ are doubles $\rho = \{ f, \varphi \}$, which is identified, in QM, with the density matrices ([Cohen1], [Sh1]). The expected value of an operator is expressed by a conventional formula for the density matrix:

$$E[A(a, a^+)] \equiv Tr(A\rho) = \langle f, A\varphi \rangle$$ 

(34)

where <.,.> is a Hilbert scalar product. In Equation (34), $A(a, a^+)$ does not depend on time explicitly but might depend on time through Heisenberg operators $a, a^+$.

Heisenberg representation for the elements of $F$ is introduced by a construction similar to Equations (17) and (18). The string of identifications runs as follows:

---

8  Because the time derivative is a shift along the trajectory of motion of Heisenberg operators, this additional term must be analogous to affine connection in differential geometry. [Dub1]

$$\langle f\, e^{-itH}, e^{itH} A e^{-itH} e^{itH} \varphi \rangle = \langle e^{itH} f, A \varphi\, e^{-itH} \rangle$$
$$\equiv Tr(A e^{itH} \rho\, e^{-itH}) \tag{35}$$

Because the operator $A(a, a^+)$ is arbitrary, we can symbolically express time evolution of $\rho$ as:

$$\rho(t) = e^{itH} \rho\, e^{-itH} \tag{36}$$

In the differential form, this equation is:

$$\frac{d}{dt}\rho = i[H, \rho] \tag{37}$$

Equation (37) can be formally rewritten in the form:

$$\frac{\partial \rho}{\partial t} = i\, \Theta\, \rho \tag{38}$$

In Equation (38), $\Theta$ is a Liouville "operator", or Liouvillian.[9] In fact, a finite-dimensional projection of the Liouvillian[10] is a 4-tensor:

$$\Theta \to \Theta_{mn,\, jk} \tag{39}$$

This finite-dimensional projection can be expressed as:

$$\Theta_{mn,\, jk} = H_{mj} \delta_{nk} - \delta_{mj} H_{kn} - \delta_{kn} V_{mj} \tag{40}$$

where $V_{mj}$ are matrix elements of the operator defined by Equation (32).[11] Because we treat Hilbert space $L$ as separable, there is little rigor lost in using formula (40) for an original state space. Technically, one pair of indexes acts on $L$ and another on $\bar{L}$. That corresponds to raising and lowering of indexes in tensor analysis but for now we shall ignore that difference.[12] Equation (35) can be formally integrated:

$$\rho = e^{it\Theta} \rho_0 \tag{41}$$

A meaning of this Equation is clarified if one imagines it as a formal expansion:

$$\rho(t) = \rho_0 + i\, t\, \Theta\, \rho_0 - t^2/2\,!\, \Theta\, \Theta\, \rho_0 + \dots \tag{42}$$

The question of convergence of these series does not arise because expression (42) is used practically only in truncated form up to a certain power of *t*. An exponential expression (41) is used only for demonstrations, such as, e.g., isometry of Hamiltonian evolution.

§14. **Probability current and velocity operator**. Our goal is to define currents, which obey continuity equations. The construct by which this is accomplished looks somewhat superficial and is subsumed by the notion of Wick ordering borrowed by the stochasticians from quantum field theory.[13] I define a new operation called "parity", which indicates order at which operator acts on the state space. Namely, $\overline{a}, \overline{a^+}$ indicate the operators acting on the expressions to the left and $\underline{a}, \underline{a^+}$ —the expressions to the right. We assume that only the expressions in which parity is "balanced" have non-zero expectations. Introduction of the divergence, velocity operator and (conserved) currents makes it possible to express the Liouvillan as a product of two factors:

$$\Theta_{mn}^{ij} = \nabla_{jm}^k v_k^n \tag{43}$$

We do not discuss here the necessary and sufficient conditions when it can be factorized. The operator acting on a density matrix:

$$\nabla \equiv \frac{1}{2\sqrt{2}\,i} \left( \overline{a} - \overline{a^+} - \underline{a} + \underline{a^+} \right) \tag{44}$$

represents divergence.[14] This definition becomes clearer in the original *p-q* representation:

$$\nabla = \frac{1}{2} \left( \overline{p} + \underline{p} \right) \tag{45}$$

Note that in presenting quantum mechanics in hydrodynamic terms, one also has to to introduce "left" and "right" velocities. [Pa1] Then, if we introduce the velocity operator as:

---

13  Compare to Di Nunno, Øksendal and Proske. ([Øk1] 2009)
14  Kolokoltsov [Kol2] identifies divergence with the projection image of the annihilation operator (Chapter 2). This definition is rigorous but makes nontransparent a connection of divergence with the continuity theorem.

$$v \equiv i \left( \frac{\delta \underline{H}}{\delta a} - \frac{\delta \overline{H}}{\delta a^+} \right) \tag{46}$$

we can define a conserved current as:

$$j = v \rho \tag{47}$$

The evolution equation for the density matrix (41) can be rewritten with the use of parity nabla as:

$$\frac{\partial \rho_k^i}{\partial t} = -\left( \nabla_{jm}^k v_k^n \right) \rho_k^m \tag{48}$$

Equation (48) is the continuity equation for our simplified algebraic formalism.

*Definition*:

The plaque measure $d\sigma_k$ is defined by the Equation:

$$d\sigma_k = \langle ., \phi_k \rangle \vec{n} \, dS$$

where $\vec{n}$ is a normal to the surface S and *dS* is a Euclidean measure.

**Lemma**.

For a sufficiently smooth oriented closed surface *A*:

$$\int_A \phi_k v_k^j \phi_j \, dS \equiv \int_A j_k \, d\sigma_k = 0 \tag{49}$$

*Proof.*

Formula (49) follows from a conventional Gauss theorem ([Galb1]) and Lagrange-Du Bois-Raymond lemma of calculus of variations [Trout1].

<u>Note</u>. The existence of the orthonormal basis of smooth functions of *L* (or *F*) is presumed for this proof. Usually these functions are presumed to be $C^l$ within *A* and continuous on its boundary. To consider singular functions (and charges) one has to define generalized derivatives using the Gelfand triple.

§15. **Gauss theorem**. Using the Liouvillian one can formulate an analog of the celebrated Gauss theorem in conventional calculus. [Galb1] So far, I do not know how to formulate Gauss' theorem using divergence and even whether its generalization is unique. However, the following equality seems to hold for a sufficiently smooth surface $A$ enclosing volume $V$ in a Euclidean space[15] at which $\phi_n = 0$ :

$$\int_V \Theta_{mn,jk} \phi_k \, dV = -\int_A \phi_k \Theta_{mn,jk} \, d\sigma_k \qquad (50)$$

**Proof**.

$\Theta_{mn,jk} \phi_k = H_{mj} \delta_{nk} \phi_k - \delta_{mj} H_{kn} \phi_n - V_{nk} \phi_k = H_{mj} \phi_n - \delta_{mj} H_{kn} \phi_n = H_{mj} \phi_n - \delta_{mj} \phi_n H_{nk} - \phi_k V_{kn}$ where, in the last equality we have used self-adjointness of the operators $H$ and $V$. The first term is identically zero on the surface $A$ by construction. Then apply Equation (49).

§16. **Continuity equation**. Starting with the continuity Equation (48):

$$\frac{\partial \rho_j^i}{\partial t} + (\nabla_{ij}^k, v_k^n) \rho_n^m = 0$$

we can rewrite it using current defined by Equations (46-48)

$$\frac{\partial \rho_k^i}{\partial t} + \nabla_{km}^i j_m = 0 \qquad (51)$$

§17. **Mean value (Lagrange, Clark-Ocone) theorem**. A stochastic process can be reconstructed exactly from its mean value and integral of its derivative. This constitutes a meaning of the Clark-Ocone theorem. [Jean1] In our case, the process is defined by its Hamiltonian. An expectation of the Hamiltonian is equal:

$$\langle \phi, H \psi \rangle = \int \langle \psi, \frac{\delta H}{\delta a} \phi \rangle \, da + \int \langle \frac{\delta H}{\delta a} \phi, \psi \rangle \, da + \lambda \langle \phi, \psi \rangle \qquad (52)$$

where $\lambda$ is an arbitrary number, which does not belong to the spectrum of $H$.

---

15 We consider Hilbert space $L$ realized as a functional space over some n-dimensional Euclidean space. This simplified setting is not necessary but we follow it anyway.

*Proof.*

First, let's notice that, because of the self-adjointness of the operator $H$, it has a complete system of eigenfuctions. Second, because Equation (53) is linear in Hamiltonian operator we can consider the functions φ and ψ belonging to the spectrum of $H$, $\sigma(H)$. Furthermore, consider an operator:

$$D = H - \lambda I$$

which is non-zero if $\lambda \notin \sigma(H)$. If we apply the delta derivative to this operator, we obtain:

$$\delta D \equiv \delta H = \frac{\delta H}{\delta a}\dot{a}\,dt + \frac{\delta H}{\delta a^+}\dot{a}^+\,dt = \frac{\delta H}{\delta a}\,d\,a + \frac{\delta H}{\delta a^+}\,d\,a^+ \tag{53}$$

Integrating Equation (53) and using the self-adjoint properties of the scalar product, we arrive at Equation (52) with an undefined integration constant instead of the original λ. Observation that

$$\langle \phi, \frac{\partial D}{\partial \lambda}\phi \rangle = \langle \phi, \frac{\partial H}{\partial \lambda}\phi \rangle - \langle \phi, \phi \rangle$$

finishes the proof.

## Appendix A. **Valuation of Portfolios with Stochastic Payoff**

This Appendix formally does not depend on the formalism of the main paper. It only demonstrates the utility of algebraic calculus of variations for one simple problem—the pricing of a portfolio with stochastic payoff.

Currently, most pricing algorithms present extensions of the Black-Scholes problem. Namely, the asset price processes could have been arbitrarily complex but the payoff function was deterministic, as in the case of call. [Hul1] However, in many problems, especially with alternative investments, the payoffs are undetermined as well. For instance, in shipping derivatives [Al1] a payoff of a given strategy depends not only on usage of ships for carrying bulk trade but also on indeterminate lay-ups and returns to active service. In our formulation, we have a portfolio of *n* assets $U_i, i = 1, \ldots n$, returns of which obey standard diffusion equations:

$$d\,U_i = a_i dt + b_i\,d\,\tilde{W}_i \tag{A.1}$$

Here, $a_i$ are expected returns, $b_i$ are the asset volatilities and $\tilde{W}_i$ are standard Brownian motions. Payoff on the portfolio of assets is nonlinear, stochastic and obeys the following equation:

$$d\,F(x) = \sum_i f_i(x)\,d\,W_i \qquad (A.2)$$

In Equation (A.2), $f_i$ are the predictable functions of their arguments and $W_i$ are standard Brownian motions. Then, the application of Itô-Kunita-Wentzel formula [Jean1, Section 1.5.3] for the infinitesimal change in the value of the portfolio provides for the following expression:

$$\begin{aligned}
dF(U) &= \nabla_{U_i} F(dU_i) + \frac{1}{2}\nabla_{U_i}\nabla_{U_j}F(dU^i)^2 \\
&+ \nabla_{U_j}f_j(dU^j\,dW_i) + f(U_i)\,dW^i = a_i\nabla_{U_i}F\,dt + b_i d\,\tilde{W}_i\nabla_{U_i}F\,dt \\
&+ \frac{1}{2}b_ib_j\nabla_{U_i}\nabla_{U_j}F\,dt + \nabla_{U_i}f_j A_i^j\,dt + f_i\,dW^i
\end{aligned} \qquad (A.3)$$

In Equation (A.3), $A_i^j\,dt = dU^j\,dW_i$ and the index of nabla operator indicates a directional derivative.

Using an arbitrary trial function $V_i$, which we assume to be predictable, we can write down a first-order optimality condition for the evolution of the portfolio:

$$\begin{aligned}
\int V_i dF(U) &= \int V_i[a_i\nabla_{U_i}F + b_i\nabla_{U_i}\dot{\tilde{W}} + \frac{1}{2}B_{ij}\nabla_{U_i}\nabla_{U_j}F]\,dt \\
+ \int V_i\nabla_{U_i}f_j A_i^j\,dt + \int V_i f_i\,dW^i &= \int V_i[a_i\nabla_{U_i}F + V_i\nabla_{U_i}f_j A_i^j + \frac{1}{2}B_{ij}\nabla_{U_i}\nabla_{U_j}F]\,dt \\
&+ \int V_i b_i\nabla_{U_i}F\,d\,\tilde{W}_i + \int V_i f_i\,dW^i = 0
\end{aligned} \qquad (A.4)$$

In Equation (A.4) $B_{ij} = b_i b_j$. Furthermore, we used a conventional shorthand: $\dot{\tilde{W}}\,dt \equiv d\,\tilde{W}$.

Now, we can use Malliavin's formula of integration by parts for the last two terms.

$$\int_0^T f_i\,dW^i = f_i(W^i(T) - W^i(0)) - \int W^i D_{W^i} f_i \qquad (A.5)$$

where $W_i$ is a Brownian motion and $D_{Wi}$ is the Malliavin derivative. We further assume that both $W^i$ and $\tilde{W}_i$ are Brownian bridges so that $W(0) = W(T)$, $\tilde{W}(0) = \tilde{W}(T)$. In the lingo of mathematical finance this means that the underlying asset prices settle at $t=T$—at the same time the entire portfolio is settled. This assumption nullifies the first term in Equation (A.5).

Using Lagrange-Du Bois-Raymond Lemma [Trout1], we can remove the integration from the Equation (A.4), which results in the SPDE of the following form:

$$[a_i \nabla_{U_i} F + \frac{1}{2} B_{ij} \nabla_{U_i} \nabla_{U_j} F + \nabla_{U_i} f_j A_i^j] dt$$
$$= \tilde{W}_i D_{\tilde{W}^i}(b_i \nabla_{U_i} F) + W_i D_{W^i} f_i(U)$$

(A.6)

Hence, Equation (A.6) acquires a standard form of diffusion equation in functional space, except for the stochastic right hand side. Finally, using the definition of divergence [Tha1], we can write the expectation of the right hand side in the form of divergence of a stochastic current:

$$[a_i \nabla_{U_i} F + \frac{1}{2} B_{ij} \nabla_{U_i} \nabla_{U_j} F + \nabla_{U_i} f_j A_i^j] = -(\hat{\nabla}_W j_1 + \hat{\nabla}_{\tilde{W}} j_2)$$

(A.7)

In Equation (A.7),

$$E^x[W D_W j_1] = E^x[\hat{\nabla} j_1]$$

(A.8)

$$E^x[\tilde{W} D_{\tilde{W}} j_2] = E^x[\hat{\nabla} j_2]$$

Probability current is given by the following expression:

$$j = j_1 + j_2 = b_i \nabla_{U_i} F + f$$

(A.9)

If $\nabla j = 0$, Equation (7) acquires a standard form of a diffusion in functional space.

<u>Note 1</u>: All computations above obviously can be rewritten in terms of creation-annihilation operators inverting arrows in Equation (1) of the main text and using the following equality:

$$a^+ a + a\, a^+ = -\Delta + q^2 \tag{A.10}$$

Equations (1), (2) and (A. 10) allow expression of the generator of diffusion in the canonical form:

$$\hat{A} = \frac{1}{2}\sigma^2 a^+ a + \lambda a + \bar{\lambda} a^+ + c \tag{A.11}$$

where σ, b, λ and c are coefficients.

<u>Note 2</u>: Conservation of current described by Equation (A.9) can be expressed in the form of a Poisson equation on the payoff function $F$ if and only if $\partial_i b \perp \partial_i f$ .

## Appendix B. **Semigroup Properties of the Hamiltonian Operator**

The problem discussed in this Appendix is a variation on the problem 1.5 from Di Nunno, Øksendal and Proske [Øk1].

Semigroup property of an operator $P^t = e^{-t\hat{H}}$ can be expressed as:

$$P^T = P^t \circ P^{T-t} \tag{B.1}$$

The following chain of equalities thus follows:

$$\begin{aligned}
\langle f(T), \varphi \rangle &= \langle f(0), \varphi \rangle + \langle f, \int_0^T \frac{\partial}{\partial t} P^{T-t} \varphi\, dt \rangle \\
&= \langle f(0), \varphi \rangle + \langle f, \int_0^T \frac{\partial}{\partial t} e^{(T-t)H} \varphi\, dt \rangle
\end{aligned} \tag{B.2}$$

and, using the Equation (52) of the main text:

$$\langle f(T), \varphi \rangle = \langle f(0), \varphi \rangle + \langle f, \int_0^T \frac{\partial}{\partial t} e^{(T-t)H} \varphi \, dt \rangle$$

$$= \langle f(0), \varphi \rangle + \langle f, \int_0^T \frac{\delta H}{\delta a} \dot{a} \, P^{T-t} \varphi \, dt \rangle + \langle f, \int_0^T \frac{\delta H}{\delta a^+} \dot{a}^+ \, P^{T-t} \varphi \, dt \rangle \qquad \text{(B.3)}$$

$$= \langle f(0), \varphi \rangle + \langle f, \int_0^T (\frac{\delta H}{\delta a} P^{T-t}(a^+, a) \, da) \varphi \rangle + \langle f, \int_0^T (\frac{\delta H}{\delta a^+} P^{T-t}(a^+, a) \, d \, a^+) \varphi \rangle$$

Comparison of right hand side with left hand side of (B.3) for T=0 provides λ=0. The last line in Equation (B.3) follows from the note in §4. We explicitly wrote the arguments of evolution operator in the last line to emphasize that the Heisenberg operators are taken at the current time.

<u>Note</u>: Compare (B.3) with the formula of Di Nunno, Øksendal and Proske in their Exercise 1.5.

Appendix C. **Approximate Measure of the High Frequency Market Making**

Carmona and Webster in the paper "High frequency market making" [Car1] derived their Equation (4.20) for the measure (in a probabilistic sense), which is placed by the market maker on the beliefs of his clients. In the setting of [Car1], an objective functional depends on the abnormal return of the stocks obeying the following set of equations:

$$d\,\alpha_t^i = -\rho\,\alpha_t^i\,dt + \sigma\,d\,M_t^i + \nu\,d\,W_t^i \qquad \text{(C.1)}$$

where ρ>0 is the rate of mean reversion, $M^i$ and $W^i$ are the Wiener processes, which are mutually independent and *i=1, 2... n*. In Equation (C.1), σ is the asset volatility and ν is the amplitude of the microstructure noise. A simplified objective functional has the form:

$$J = \frac{1}{n} Tr\left[V^i \alpha_j f\left(\vec{\alpha}\right)\right] \tag{C.2}$$

where $f$ is a sufficiently smooth <u>scalar</u> function of its argument.

Applying the definition of $B_{ij}$ after Equation (A.4), we have:

$$\frac{1}{n} Tr\left[B_{ij}\right] = \sigma^2 + \nu^2 \tag{C.2}$$

Other replacements in Appendix A acquire the following form:

$$F \rightarrow f \cdot Id \,,$$
$$f_i \rightarrow \nu \delta_{ij} \,,$$
$$A^i_j \rightarrow -\rho \, \delta_{ij}$$
$$\tilde{W}_i \rightarrow W_i$$

Equation (A.4) now reads as:

$$\int V_i \, df = \int V_i \left[-\rho \, Id \, \nabla f + \frac{1}{2}(\sigma^2 + \nu^2)\Delta f\right] dt$$
$$+ \int V_i \nu \nabla_i f \, d W_i \tag{C.3}$$

which coincides with Equation (4.20) in Carmona and Webster [Car1].